\documentclass
[aps,amsmath,amssymb,superscriptaddress]{revtex4}%
\usepackage{ulem}
\usepackage{amsfonts}
\usepackage{amsmath}
\usepackage{amssymb}
\usepackage{graphicx}
\usepackage{textcomp}%
\providecommand{\U}[1]{\protect\rule{.1in}{.1in}}

\begin{document}

\title{Temporal solitons in optical microresonators}
\author{T.~Herr}
\affiliation{\'{E}cole Polytechnique F\'{e}d\'{e}rale de Lausanne (EPFL), 1015,
Lausanne, Switzerland}
\author{V.~Brasch}
\affiliation{\'{E}cole Polytechnique F\'{e}d\'{e}rale de Lausanne (EPFL), 1015,
Lausanne, Switzerland}
\author{J.D.~Jost}
\affiliation{\'{E}cole Polytechnique F\'{e}d\'{e}rale de Lausanne (EPFL), 1015,
Lausanne, Switzerland}
\author{C.Y.~Wang}
\affiliation{\'{E}cole Polytechnique F\'{e}d\'{e}rale de Lausanne (EPFL), 1015,
Lausanne, Switzerland}
\author{N.M.~Kondratiev}
\affiliation{Faculty of Physics, M.V.Lomonosov Moscow State University, Moscow 119991, Russia}
\author{M.~L.~Gorodetsky}
\email{michael.gorodetsky@gmail.com}
\affiliation{Faculty of Physics, M.V.Lomonosov Moscow State University, Moscow 119991, Russia}
\affiliation{Russian Quantum Center, Skolkovo 143025, Russia} 
\author{T.~J.~Kippenberg}
\email{tobias.kippenberg@epfl.ch}
\affiliation{\'{E}cole Polytechnique F\'{e}d\'{e}rale de Lausanne (EPFL), 1015, Lausanne, Switzerland}

\begin{abstract}
\textbf{Dissipative solitons can emerge in a wide variety of dissipative nonlinear systems throughout the fields of optics, medicine or biology\cite{Akhmediev2008}. Their dissipative nature requires them to be continuously in contact with an external energy source, and renders their attractor highly stable and robust against fluctuations\cite{Akhmediev2008,Grelu2012}. Dissipative solitons can also exist in Kerr-nonlinear optical resonators and rely on the double balance between parametric gain and resonator loss on the one hand and nonlinearity and diffraction or dispersion on the other hand. Mathematically these solitons are solution to the Lugiato-Lefever equation\cite{Lugiato1987} and exist on top of a continuous wave (cw) background\cite{Wabnitz1993,Firth2002}. While spatial dissipative solitons in optical resonators\cite{Segev1998,Firth2002} have been widely studied, their temporal counter-part\cite{Wabnitz1993,Firth2002} is difficult to access and has only recently been generated in laser driven optical fiber-loops, using an additional writing laser\cite{Leo2010}. Here we report the observation of temporal dissipative solitons in a high-Q optical microresonator. The solitons are spontaneously generated when the pump laser is tuned through the effective zero detuning point of a high-Q resonance, leading to an effective red-detuned pumping. Red-detuned pumping marks a fundamentally new operating regime in nonlinear microresonators. While usually unstable\cite{Carmon2004} this regime acquires unique stability in the presence of solitons without any active feedback on the system. The number of solitons in the resonator can be controlled via the pump laser detuning and transitions to and between soliton states are associated with discontinuous steps in the resonator transmission. 
Beyond enabling to study soliton physics such as soliton crystals\cite{Akhmediev2008,Grelu2012} our observations open the route towards compact, high repetition-rate femto-second sources, where the operating wavelength is not bound to the availability of broadband laser gain media. The single soliton states correspond in the frequency domain to low-noise optical frequency combs with smooth spectral envelopes, critical to applications in broadband spectroscopy, telecommunications, astronomy and low phase-noise microwave generation.}
\end{abstract}
\maketitle

High-Q, nonlinear optical microresonators (more precisely: dielectric whispering gallery mode or ring-type resonators) have recently attracted growing attention in the scientific community. In particular frequency comb generation in high-Q microresonators has, within a few years, evolved to a research field of its own\cite{DelHaye2007,Savchenkov2008c,Levy2010,Razzari2010,Kippenberg2011,Papp2012,Li2012}. In microresonator based frequency combs a cw pump laser is coupled to a high finesse resonator(whereby, for thermal stability reasons, the pump laser is effectively blue detuned; see below). The high light intensities resulting from the high cavity finesse and the strong modal confinement enable cascaded four-wave mixing (FWM), which can give rise to hundreds of equidistant and coherent optical lines that, together with the pump laser, can constitute a frequency comb (cf. Fig.~1c). The comb line spacing corresponds to the free-spectral range of the microresonator or equivalently the inverse cavity roundtrip time $T_R$. The achievable comb line spacings range from several GHz up to THz frequencies. FWM based microresonator combs can perform on a level required for optical frequency metrology applications\cite{DelHaye2008,Papp2012,DelHaye2012}. However, these systems often suffer from significant frequency and amplitude noise\cite{Herr2012} and, unlike conventional mode-locked laser based frequency combs, do not correspond to ultra-short pulses in the time domain. The latter can be understood by the constant but arbitrary phase relations between the comb lines, which result from the formation process (cf. Fig.~1c)\cite{Herr2012}. External line-by-line phase and intensity adjustment may be used after comb generation for pulse-shaping\cite{Ferdous2011,Papp2011a}, but this is restricted to only a small number of comb modes. We note, that very recently evidence for direct ultra-short pulse generation in a Si$_3$N$_4$ microresonator has been found, which is to date unexplained\cite{Saha2012a}. In a different system comprising a fiber cavity with laser gain and a nonlinear high-index silica microresonator in filter configuration, generation of high-repetition rate pico-second pulses was shown\cite{Peccianti2012}.

In the case of a strongly driven nonlinear microresonator, the intracavity field as function of the pump laser detuning cannot be described by a Lorentzian-shape resonance (cf. Fig.~1c). Instead, the resonance is asymmetrically shifted towards lower frequencies by the Kerr-nonlinearity when the intracavity power increases. This leads to a bistable behavior, that is, two possible solutions for the intracavity power can exist for a particular pump laser detuning. The two solutions are usually referred to as {\it upper} branch (higher intracavity power) and {\it lower} branch (lower intracavity power) solutions (cf. Fig.~1c). These two branches correspond respectively to blue and red detuned operation of the pump laser. Besides the Kerr-nonlinear resonance shift, an increased intracavity power also results in an additional shift of the resonance frequency towards lower frequencies via a combined effect of thermal expansion and thermal refractive index change (induced by absorptive heating). The combined  Kerr-nonlinear and thermal effects lead to an non-Lorentzian, triangular resonance shape ({\it thermal triangle}) when the pump laser is scanned with decreasing optical frequency over the resonance (cf. Fig 2a, inset)\cite{Ilchenko1992a,Carmon2004}. It is important to note that the resonance frequency self-locks to the pump laser\cite{Carmon2004}, when the pump laser is blue detuned (upper branch) with respect to the effective resonance frequency; the system is thermally unstable if the pump laser is effectively red detuned (lower branch)\cite{Ilchenko1992a,Carmon2004}. This self-stability is exploited in microresonator based frequency comb generation where the pump laser is operated effectively blue detuned.

In this work, we show that tuning the pump laser through the effective zero detuning frequency, into the lower branch (effectively red detuned) after previously following the upper branch (effectively blue detuned; observing concomitant FWM), leads to the formation of temporal dissipative cavity solitons in a microresonator. This regime is qualitatively different from the stable operating regime of microresonator based frequency combs used so far, which have relied on pumping the resonator from the blue sideband and have not crossed the zero-detuning point, after which the resonator becomes thermally unstable. In contrast to fiber cavity experiments\cite{Leo2010}, the soliton pulses form spontaneously without the need for external stimulation. The number of solitons formed can be controlled by the pump laser detuning. The generated solitons remain stable until the pump laser is switched off without the need for active feedback on either the resonator or the pump laser. This remarkable stability in the presence of solitons, despite operating on the usually thermally unstable lower branch (effectively red detuned pump laser), will be discussed below. Our discovery enables converting a cw laser into a train of femto-second pulses, which corresponds to a low noise smooth spectral envelope frequency comb in the time domain.

In the present work we use a MgF$_2$ microresonator\cite{Grudinin2006,Liang2011,Grudinin2012,Wang2013} that meets the basic requirements for temporal dissipative soliton formation, that is Kerr-nonlinearity (also responsible for the parametric gain) and anomalous dispersion. The microresonator is characterized by a free-spectral range (FSR) of 35.2~GHz (Fig. 1a and Methods) and a coupled resonance width of 450~kHz. The resonator's measured (cf. Fig.~1b)\cite{DelHaye2009a} anomalous group velocity dispersion (GVD) is $\beta_2=-9.39$~ps$^2$km$^{-1}$, which in the context of microresonators can be conveniently expressed in terms of the parameter\cite{Savchenkov2011,Herr2012} $D_2 =-c/n_0\cdot D_1^2\cdot\beta_2=2\pi \cdot 16$~kHz that describes the deviation of the resonance frequencies $\omega_{\mu}=\omega_{0}+D_{1}\mu +\tfrac{1}{2}D_2\mu^{2} +\tfrac{1}{6}D_3\mu^{3} + ...$ from an equidistant frequency grid defined by $\omega_0+\mu D_1$, where $c$ is the speed of light and $n_0$ the refractive index of MgF$_2$ and $\mu$ the relative mode number with respect to the pumped mode $\omega_0$, where $\mu=0$ by definition. $D_{1}/2\pi$ is the FSR of the resonator at the frequency $\omega_0$. The frequency deviation increases quadratically with increasing relative mode number $\mu$ as evidenced in Fig.~1b; $D_3$ and higher order terms can be neglected in the present case.

To search for soliton states in the microresonator we scan a pump laser (fiber laser; wavelength $1553$~nm; linewidth~$\sim10$~kHz) with decreasing optical frequency $\omega_{p}$ over a high-$Q$ resonance of the crystalline MgF$_{2}$ resonator. This approach is motivated by the pump laser detuning being a critical parameter for the existence of cavity solitons\cite{Wabnitz1993,Barashenkov1996,Leo2010}. Fig.~2b shows the evolution of the optical spectrum during the laser scan. Reducing the laser-cavity detuning leads to a build-up of intra-cavity
power and once a critical power threshold\cite{Kippenberg2004a,Matsko2005a} is reached widely spaced primary sidebands are generated by FWM, followed by secondary lines filling in the spectral gaps as frequently observed in FWM based microresonator combs\cite{Ferdous2011,Herr2012,Grudinin2012}. While performing the scan, the RF (radio frequency) signal (electronically down-mixed to 20 $\mathrm{MHz}$) that results from the beating between neighboring comb lines is sampled and Fourier-transformed. A necessary signature of stable soliton formation is a low-noise, narrow RF signal, resulting from the repetitive output-coupling of a soliton pulse. The Fourier-transformed, sampled RF signal is contained in Fig. 2c. Indeed, we observe a transition from a broad, noisy RF signal to a single, low-noise RF beat note for a particular laser detuning. This transition coincides with the beginning of a series of discrete steps in the transmission, which deviates markedly from the expected thermal triangle (the RF beatnote remains narrow throughout all steps). Note that, observations similar to the discrete transmission steps have been made in a $\chi^{(2)}$ nonlinear microwave resonator, and connected to soliton formation\cite{Gasch1984}. To determine the effective pump laser detuning, we record a Pound-Drever-Hall (PDH) error signal\cite{Drever1983} while scanning over the resonance. Strinkingly, the first step, that is the transition to low noise, coincides with the zero crossing of the PDH signal, which marks the effective zero detuning frequency. This observation implies that the occurrence of the steps coincides with the transition from the upper branch regime of microresonator based FWM combs to the so far unexplored lower branch regime.

To understand the intriguing observations of discrete steps in the transmission and the possible connection to soliton-formation we carry out numerical simulations based on the coupled mode-equations approach (cf. Methods)\cite{Chembo2010}. Note that the coupled mode equations are equivalent to the Lugiato-Lefever equation when third and higher order dispersion can be neglected\cite{Matsko2011a}, which is the case in the present microresonator (cf. Fig. 1b). The simulated system corresponds to a typical MgF$_{2}$ microresonator similar to the one used for the experiments. The remaining resonator parameters are refractive and nonlinear indices, as well as the effective mode-volume. We neglect effects of non-unity mode-overlap, interactions with other mode families, any particularities of the resonator geometry and thermal effects. The resulting set of coupled mode equations is numerically propagated in time (cf. Methods and Supplementary Information, SI). Results of a numerical simulation including 101 comb modes are shown in Fig.~3. The blue curve in Fig. 3a shows the simulated intracavity power as function of the normalized detuning $\zeta_{o}=2(\omega_{0}-\omega_{p})/\kappa$ of the pump laser $\omega_\mathrm{p}$ from the cold resonance frequency, where $\kappa=\omega_{0}/Q$ denotes the cavity decay rate. Note that the intracavity power (the blue curve in Fig.~3a) is equivalent to the experimental transmission trace in Fig. 2a (an increased transmission corresponds to a drop in intracavity power).  Due to the nonlinear Kerr-shift of the resonance frequency, the effective detuning between pump laser and resonance is smaller than $\zeta_{0}$. Remarkably, the step features are very well reproduced, implying that the simulation includes all relevant physical mechanism. In agreement with the experiment the number and height of steps fluctuate in repeated numerical scans as a result of random seeding of the optical modes (corresponding to vacuum fluctuations). Numerically tracing out all possible comb evolutions yields the orange curves in Fig. 3a. The first part of the evolution of the optical spectrum, shown in Fig.~3b follows the known pathway for FWM based comb formation\cite{Herr2012}. Later on, with each step in the transmission the optical spectrum becomes less modulated until it eventually reaches a perfectly smooth envelope state (frame XI).

To reveal the potential underlying soliton formation we investigate the time dependent waveform in Fig.~3c by phase-coherently adding the individual simulated optical modes. Indeed, the first step (frame V) corresponds to a transition to a state where multiple pulses inside the cavity exist. Further steps can be associated with a stepwise reduction of the number of pulses propagating in the resonator. The separation between multiple pulses in the resonator is random.  To confirm the soliton nature of these pulses we perform a simulation of 501 modes (cf. Fig. 3d,e,f) and analyze a state of five pulses. We compare the numerical simulation with an approximate analytical solution of the Lugiato-Lefever equation\cite{Lugiato1987}. For multiple solitons the analytical solution \cite{Wabnitz1993} has the form
\begin{equation}
\ensuremath{\Psi(\phi)\simeq C_{1}+C_{2}\cdot\sum_{j=1}^{N}{\rm sech}(\sqrt{\frac{2(\omega_0-\omega_p)}{D_{2}}}(\phi-\phi_{j}))}
\label{eq:AnalyticalSoliton}
\end{equation}
where $\Psi$ denotes the complex field amplitude, $\phi$ the angular
coordinate inside the resonator, $\phi_{j}$ the angular coordinate
of the $j$th soliton, $\omega_{p}$ the pump frequency, and where $N$
is the number of solitons. The complex numbers $C_{1}$ and $C_{2}$ are fully determined by the resonator parameters and the pump conditions and the ratio $|C_2|^2$/$|C_1|^2$ of soliton peak power to cw background can typically exceed several hundreds (cf. SI for details). Indeed, the close to perfect match between analytic solution and numerical result shows that the pulses forming in the microresonator are stable temporal dissipative cavity solitons. These solitons emerge from the modulated intracavity waveform, which may explain their spontaneous formation as opposed to fiber cavities where stimulating writing pulses are required\cite{Leo2010}. Eq. \ref{eq:AnalyticalSoliton} allows for the estimation of the minimal temporal soliton width (FWHM)
\begin{align}
 \Delta t_\mathrm{min}^\mathrm{FWHM} \approx 2 \sqrt{\frac{-\beta_2}{\gamma {\cal F} P_\mathrm{in}}},\label{eq:PulseDuration}
 \end{align}
where ${\cal F}$ is the cavity's finesse, $P_\mathrm{in}$ the coupled pump power, $\beta_2$ the GVD and $\gamma=\tfrac{\omega}{c}\tfrac{n_2}{A_\mathrm{eff}}$ the effective nonlinearity with the nonlinear mode area $A_\mathrm{eff}$ and the nonlinear refractive index $n_2$ (cf. SI).

Having shown the soliton nature of the pulses in numerical simulations, we can interpret the blue curve in Fig.~3a based on general limits\cite{Barashenkov1996} applying to solitons as solutions of the Lugiato-Lefever equation. Adopting these criteria for the present case we identify three main regions in Fig.~3a colored red, yellow and green (cf. SI). Solitons with a constant temporal envelope can only exist in the green area. While the yellow area still allows for solitons with a time varying envelope (``breather solitons'' \cite{Matsko2012}), solitons can not exist in the red area. Note that in the red area on the left, the system may undergo chaotic Hopf-bifurcations\cite{Barashenkov1996}. 

For different number of solitons we can derive the total power inside the resonator by averaging the respective analytic soliton solution (eq. \ref{eq:AnalyticalSoliton}) over one cavity roundtrip time. The result is shown as dark gray dashed lines in Fig. 3a, and is in excellent agreement with the numerically observed steps (to account for the limitation due to the low mode number an additional correction factor of order unity is applied). The intracavity power changes discontinuously with the number of solitons present in the cavity (cf. SI).

To experimentally generate the soliton states, we develop a method that allows for reliably tuning into the desired soliton state. This method relies on tuning the laser with an appropriately chosen tuning speed into the soliton state (see SI for detail). Once generated, the solitons remain stable until the pump laser is switched off and no active stabilization or feedback on either the resonator or the pump laser is required. The latter observation is surprising as pumping a microresonator on the lower branch (effectively red detuned) implies thermal instability\cite{Ilchenko1992a,Carmon2004} and would usually require active stabilization techniques. In the presence of solitons, however, the fraction of the pump light that propagates at a similar velocity together with the high intensity soliton inside the resonator, experiences a much larger phase shift in one cavity roundtrip. While the main portion of the pump light is effectively red detuned (lower branch) the small portion overlapping with the soliton inside the resonator is effectively blue detuned (upper branch) and responsible for the self-stability of the system\cite{Carmon2004}. This is evidenced by the series of steps that correspond to a set of small thermal triangles (characteristic for stable blue detuning; one triangle per realized soliton state) as shown in Fig.~2a (experiment) and Fig.~3a (simulation). We note that this self-stability is unique to micro-resonators and not observed in fiber cavities. A more detailed discussion can be found in the SI. For clarity we note that a hypothetical disturbance in the laser detuning larger than the length of the small thermal triangles would terminate the soliton operation. 

Having access to stable operation of these soliton states, we experimentally investigate - in addition to their RF beatnote and optical spectrum (Fig.~4a) - their temporal characteristics by performing a frequency-resolved optical gating experiment (FROG, Fig.~4b)\cite{Kane1993,Dudley1999}. This corresponds to a second-harmonic generation (SHG) autocorrelation experiment, where the frequency-doubled light is spectrally resolved
(Fig.~5b and Methods). In contrast to auto-correlation, the FROG method allows a reliable identification of ultra-short pulses via the associated minimal bandwidth of the SHG spectrum given by the time-bandwidth-product (TBP). A comparison of auto-correlation and FROG method is provided in the SI.

In full consistency with the numerical simulations,
we observe single and multiple soliton states. The
single soliton state is characterized by a smooth spectral envelope, without
spectral gaps. The power spectral envelope exhibits a sech$^{2}$-shape (3~dB bandwidth 1.6~THz corresponding to more than 45~modes) as expected from the Fourier transform of a sech-shaped soliton pulse. Based on the TBP for solitons of~0.315 (cf. SI) the expected pulse duration is 197~fs. The observed low phase noise RF beatnote is resolution bandwidth limited
to 1 kHz and its signal-to-noise ratio exceeds 60 dB. The FROG trace shows a train of pulses well separated by the cavity roundtrip time of $T_{R}=$28.4~ps, corresponding to the FSR of 35.2~GHz. The multi-soliton states (here
shown for the case of two and five solitons), show a more structured
optical spectrum. This structure reflects the number and distribution
of solitons in the cavity (cf. SI). The RF beat note generated in the multi-soliton
states is of similar quality as in the single soliton state. Importantly,
the FROG measurement allows for a full reconstruction (neglecting
a time direction ambiguity) of intensity and phase of the pulses (cf.
Fig. 5a). The reconstructed intensity profile is consistent with the expected
sech$^{2}$-shape for solitons and the reconstructed temporal width of 200 fs (FWHM) is in agreement with the bandwidth of the optical spectrum and the expectation based on eq. (\ref{eq:PulseDuration}). The FROG traces show that it is the full spectrum that contributes to each pulse.

To further corroborate the presented results an independent intensity sampling method is applied to a single soliton state. The high repetition rate prohibits a direct sampling, which would require hundreds of GHz bandwidth in detection and recording. This limitation can be overcome by stretching the optical waveform in time using an Ultrafast Temporal Magnifier\cite{Salem2009} (PicoLuz LLC, Fig.~5d). While the time resolution of about 2.5 ps does not allow for determining the duration of the pulse, this single-shot method, as opposed to auto-correlation-type experiments, does not rely on averaging. The results in Fig.~5c clearly show optical pulses separated by $T_{R}$, with constant pulse amplitudes, as expected for solitons. We emphasize that in all temporal characterization experiments neither phase and intensity adjustment (except for pump suppression) nor spectral filtering are applied. 
The gain window ($\geq 4$~THz) of the optical amplifier used before temporal characterization supports more than 100 comb modes.

Combining experimental, numerical and analytical results, we have demonstrated spontaneous formation of dissipative temporal cavity solitons in a MgF$_{2}$ microresonator. The duration of these soliton pulses depends on the pump power and the dispersion of the resonator (cf. eq.~2). In the present case the optical pulses are in the range from 200~fs. Given the possibility of dispersion engineering in microresonators and the broadband nature of the parametric gain, significantly shorter pulses are conceivable. The stable solitons allow ultra-short pulses to be continuously coupled out of the microresonator yielding a train of ultra-short pulses. If only one soliton is present in the cavity this pulse train corresponds to a low-noise, optical frequency comb with only low line-to-line power variation. Comparable frequency combs have so far not been generated in microresonators. As our results only depend on the generic properties of Kerr-nonlinear microresonators with anomalous dispersion, they apply equally to other microresonator comb platforms. We note that the observations of the step signature is not a spectrally local particularity when driving one particular resonance but is observed when driving any resonance of the same mode family within the tuning range ($\pm$0.5~nm) of the pump laser. Soliton formation, as revealed in our work, may also, at least partially, explain the generation of femto-second pulses in Si$_3$N$_4$ resonators\cite{Saha2012a} reported recently. Moreover, our results are in agreement with very recent numerical work on temporal dissipative soliton formation in microresonators\cite{Coen2013,Coen2013c,Chembo2013}.

In contrast to mode-locked lasers, which rely on laser gain, no additional element, such as a saturable absorber is required for stable operation in the microresonator case, which relies on parametric gain (A detailed discussion on the difference to mode-locked laser is contained in the SI). It is moreover worthwhile emphasizing that temporal dissipative cavity solitons are different from dissipative solitons in lasers, which already find widespread use\cite{Grelu2012}. 

From a frequency domain perspective, soliton formation enables microresonator based frequency combs with low noise and smooth spectral envelopes. These low noise spectra with unprecedentedly small line-to-line power variations are essential to frequency domain application in telecommunications\cite{Pfeifle2012}, broadband spectroscopy\cite{Diddams2007} and astronomy\cite{Steinmetz2008,Li2008a}. Recent theoretical work suggests that octave spanning combs may be obtainable in the soliton regime\cite{Coen2013,Lamont2013} directly from a microresonator. From a time domain perspective soliton formation in a microresonator enables ultra-short pulse generation in a microresonator. In combination with chip-scale\cite{Levy2010,Razzari2010} integration this opens the route towards compact, stable and low cost ultra-short pulse sources\cite{Grelu2012}, which can also operate in wavelength regimes (such as the mid-infrared\cite{Wang2013}), where broadband laser gain media do not exist. Moreover, femto-second pulses in conjunction with external broadening\cite{Dudley2006} (see SI for a first demonstration) provide a viable route to a microresonator RF-to-optical link\cite{Telle1999,Cundiff2003}. Moreover, the unique stability of dissipative solitons\cite{Grelu2012} is of interest to low phase-noise microwave generation\cite{Savchenkov2008c}.

\section*{Methods}
\subsection*{Experimental setup and parameters:}
The pump laser (fiber laser Koheras Adjustik; 1553~nm wavelength; shortterm linewidth $10$~kHz) is amplified by an erbium doped fiber amplifer (EDFA) and  evanescently coupled to the MgF$_2$ resonator (free spectral range $35.2$~GHz; refractive index $n_0=1.37$) via a tapered optical fiber. The coupled resonance width of $\kappa/2\pi=450$ kHz (quality factor $Q=4\times10^{8}$, Finesse ${\cal F} = 78 \times 10^4$) has been measured using modulation sidebands of a scanning laser. The dispersion of the resonator has been determined following ref. \cite{DelHaye2009a} to $D_2/2\pi=16$~kHz that is $\beta_2=-9.39$~ps$^2$km$^{-1}$. A typical coupled pump power of $P_\mathrm{in}=5-30$~mW leads to circulating powers $P_\mathrm{circ}$ of several hundreds of Watts. The estimated effective mode area $A_\mathrm{eff}=90 \times 10^{-12}$m$^2$ (effective mode volume $V_\mathrm{eff}=5.6\times10^{-13}\mathrm{~m}^{3}$) and the nonlinear refractive index
$n_{2}=0.9\times10^{-20}\mathrm{~m}^{2}\mathrm{W}^{-1}$ yields an estimated nonlinear parameter of $\gamma = \tfrac{\omega_\mathrm{p}}{c}\tfrac{n_2}{A_\mathrm{eff}}=4.1 \times 10^{-4}$m$^{-1}$W$^{-1}$.

\subsection*{Numerical simulation:}
The simulations are based on the coupled mode equations (cf. SI) which are numerically propagated in time using an adaptive step-size Runge-Kutta integrator. The simulated resonator is defined similar to the experimental resonator by its resonance frequencies $\omega_{\mu}$ given by $D_{1}/2\pi=35.2$~GHz, $D_{2}/2\pi=10$~kHz, $D_{3}/2\pi=-130$~Hz (these values where measured for a resonator similar to the one used here) and a quality factor of $Q=2\times10^{8}$. All other parameters are equal to the values listed for the experimental resonator. The coupled pump power is set to $P_{\mathrm{in}}=100\mathrm{~mW}$ at a pump frequency of $\omega_{p}/2\pi=193$ THz. Short, simulated pump power drops can be used to induce transitions between different soliton states.

\subsection*{FROG Experiment:}
Prior to the FROG\cite{Kane1993,Dudley1999} experiment the optical spectra are sent through
a fiber-Bragg grating for pump supression ($-30$ dB) and are subsequently
amplified to 50 mW. Dispersion compensating fiber (Thorlabs DCF3,DCF38) is
used for approximate dispersion compensation. In the FROG setup (cf.
Fig.~5b) the generated optical pulses, are interferometrically split
and recombined with a variable delay in a nonlinear BBO crystal.
This results in second harmonic generation (SHG) whenever the optical pulses in the two arms
of the interferometer overlap temporally in the BBO crystal.
The generated SHG light is spectrally resolved and recorded as a function
of delay by a CCD-spectrometer, yielding a FROG trace.
Each FROG trace consists of nearly 1000 spectra with individual exposure times of 800 ms.
The N-by-N (N=63) FROG trace of the single pulse state in Fig.~5a is analyzed using a principal component generalized projection
algorithm\cite{Kane2008}, after noise removal via Fourier-filtering.
The FROG reconstruction error is defined as $\epsilon=\frac{1}{N}\sqrt{\sum_{i,j}^{N}(M_{ij}^\mathrm{meas}-M_{ij}^\mathrm{reco})^{2}}$, where $M_{ij}^\mathrm{meas}$ and $M_{ij}^\mathrm{reco}$ denote the elements of the $N\times N$ matrices representing the measured and reconstructed FROG traces.

\section*{Acknowledgments}
The authors thank R. Salem and A. Gaeta for providing the PicoLuz, LLC Ultrafast Temporal Magnifier and advice when evaluating the data. 
This work was supported by the DARPA program QuASAR, the Swiss National Science Foundation (SFN) as well as a Marie Curie IAPP program. MLG acknowledges support from the RFBR grant 13-02-00271. The research leading to these results has received funding from the European Union Seventh Framework Programme (FP7/2007-2013) under grant agreement no. 263500.

\section*{Author Contribution}
TH designed and performed the experiments and analyzed the data. MLG and TH performed the numerical simulations. MLG developed the analytic description. VB assisted in the experiments. JDJ assisted in the Temporal Magnifier experiment. TH and MLG fabricated the sample. CYW assisted in sample fabrication. NMK assisted in developing the analytic description. TH, MLG, and TJK wrote the manuscript. TJK supervised the project.

\newpage

\begin{figure}[htbp]
\centering
\includegraphics[width=1\textwidth]{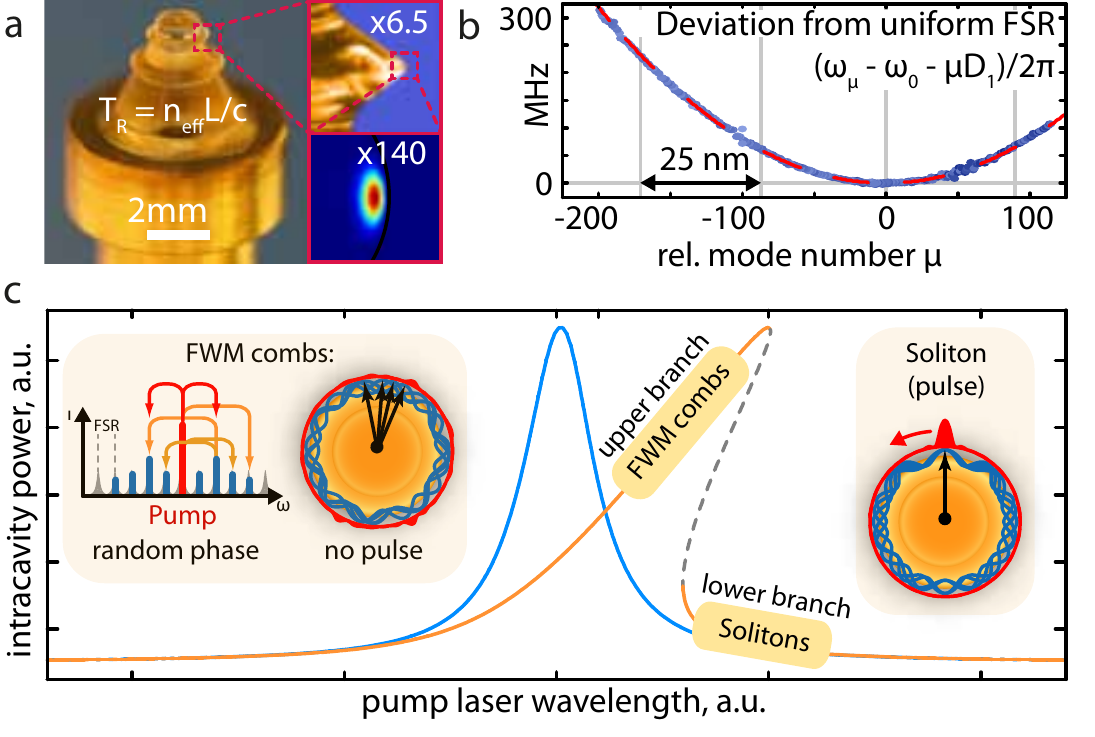}
\caption{\textbf{MgF$_2$ microresonator, dispersion and bistability. a.} MgF$_{2}$ crystal carrying two whispering-gallery-mode microresonators of different size (the smaller one with an FSR of 35.2 GHz is
used). A optical whispering-gallery-mode propagates along the circumference $L$ of the resonator within the roundtrip time $T_{R}$. The smaller panels show a magnified view of the resonator and the simulated optical mode profile. \textbf{b.} Second order anomalous dispersion (FSR increases with mode number) of the microresonator with 35.2 GHz FSR shown as the deviation of the measured resonance frequencies (blue dots) $\omega_\mu = \omega_0+ D_1  \mu+\tfrac{1}{2}D_2\mu^2 + ... $ from an equidistant frequency grid $\omega_0+\mu D_1$ (horizontal grey line), where $\mu$ denotes the relative mode number and $D_1$ corresponds to the FSR at the frequency $\omega_0$. The resonator's anomalous dispersion is accurately described by $D_2/2\pi=16$~kHz and higher order terms neglected (red dashed line). The grey vertical lines mark spectral intervals of 25~nm width ($\mu=0$ corresponds to 1553~nm).
\textbf{ c.} Bistable intracavity power as function of laser detuning for a linear cavity (blue) and a nonlinear cavity (orange). The dashed line marks the unstable branch. The regimes of FWM based microresonator combs on the upper branch and solitons on the lower branch are marked. In FWM combs the phases of the comb lines are constant but random, leading to a periodic but un-pulsed intracavity waveform (cf. left inset). The presence of a soliton implies synchronized phases and a pulsed intracavity waveform (cf. right inset).}
\end{figure}

\begin{figure}[htbp]
\centering
\includegraphics[width=0.73\textwidth]{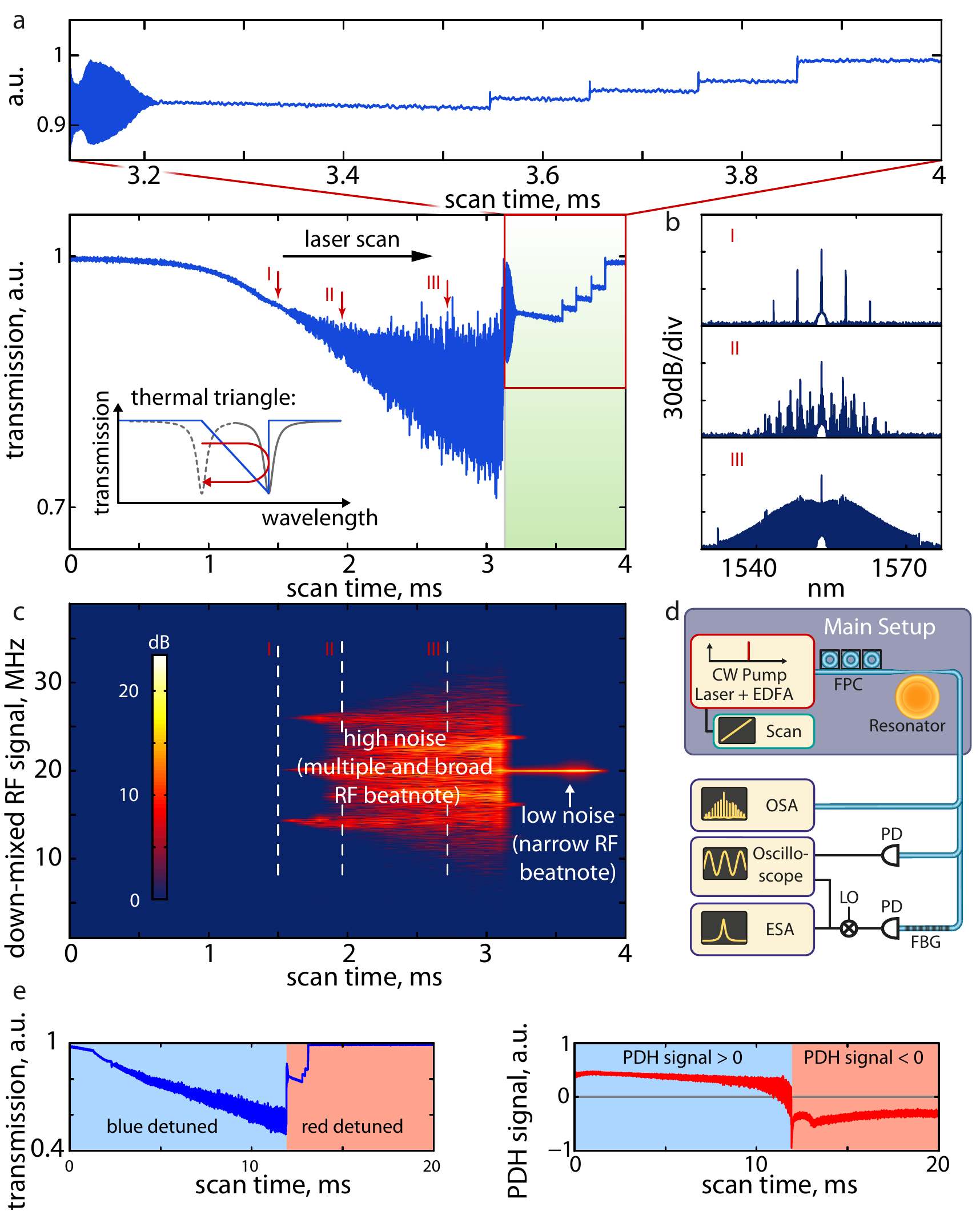}
\caption{\textbf{Transmission and beatnote a.} Transmission observed
when scanning a laser over a high-Q Kerr-nonlinear resonance in a
MgF$_{2}$ resonator (coupled pump power 5 mW). The transmission signal
follows the expected thermal triangle (cf. inset) with deviations
in the form of discrete steps (green shading). \textbf{b.} Evolution
of the optical power spectrum for three different positions in the scan; the spectrum (II) and in particular the mesa shaped one (III) exhibit a high-noise RF beat signal. \textbf{c.} Down-mixed RF beat signal.
\textbf{d.} Experimental main setup composed of pump laser and resonator
followed by an optical spectrum analyzer (OSA), an oscilloscope to record the transmission and to
sample the down-mixed beatnote (via the third harmonic of a local
oscillator LO at 11.7 GHz), and an electrical spectrum analyzer (ESA)
to monitor the beatnote. Before beatnote detection the pump is filtered
out by a narrow fiber-Bragg grating (FBG) in transmission; FPC: Fiber
polarization controller, EDFA: erbium-doped fiber
amplifier, PD: photodetector. \textbf{e.} Transmission and Pound-Drever-Hall (PDH) error signal. Effective blue/red detunings are shaded blue/red.}
\end{figure}

\begin{figure}[htbp]
\centering
\includegraphics[width=1\textwidth]{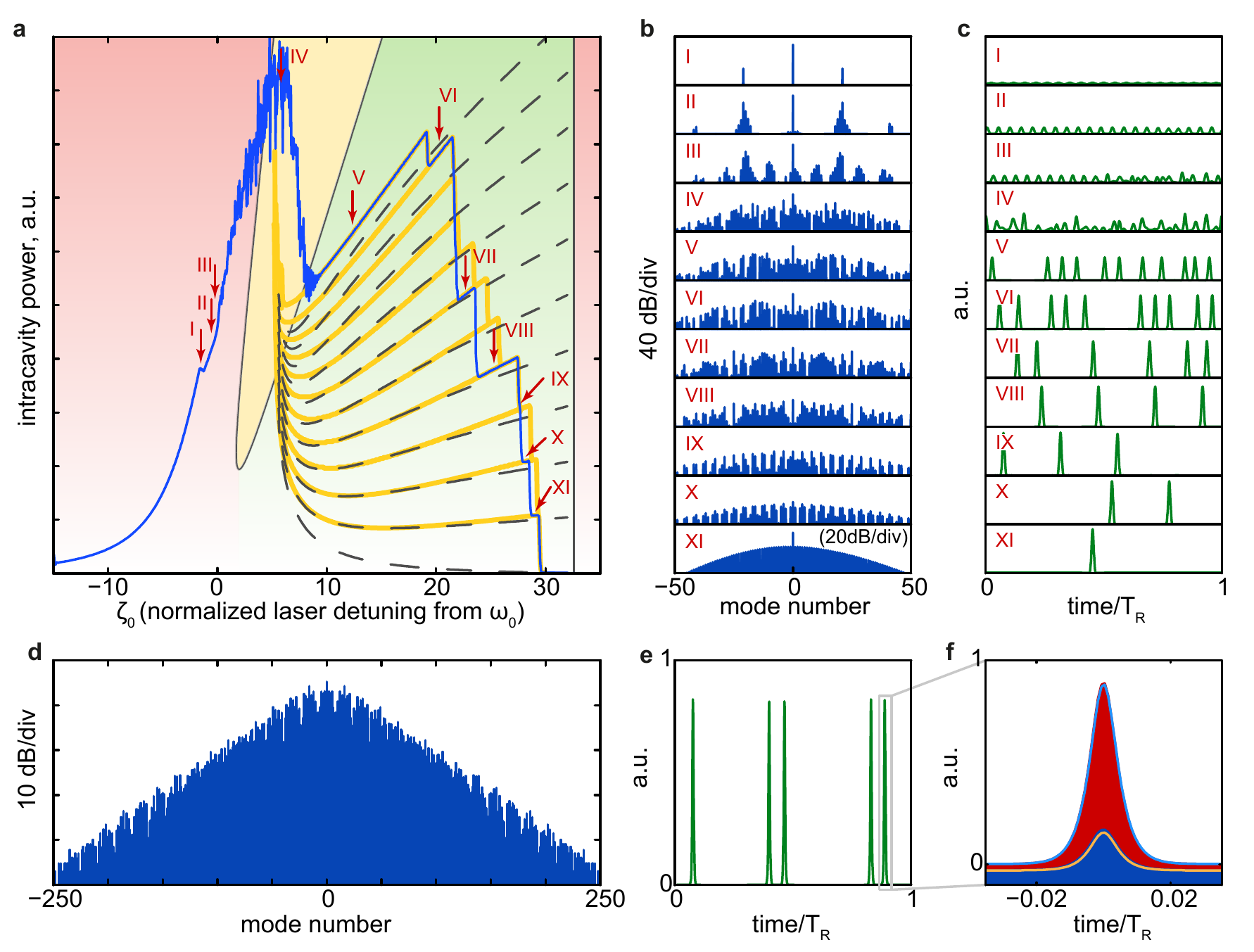}
\caption{\textbf{Numerical Simulations of dissipative temporal soliton formation in a microresonator a.} Intracavity power (blue, corresponding to the transmission signal in Fig. 2a when mirrored horizontally) during a simulated laser scan (101 simulated modes) over a resonance in a MgF$_{2}$ resonator. The step features are well reproduced. The orange lines
trace out all possible evolutions of the system during the scan. The
dashed lines show an analytical description of the steps. The green
area corresponds to the area where solitons exist, the yellow area
allows for solitons with a time variable envelope; no solitons can
exist in the red area. \textbf{b/c.} Optical spectrum and intracavity
intensity for different positions I-XI in the laser scan. \textbf{d.}
Optical spectrum obtained when simulating 501 modes and stopping the
simulated laser scan in the soliton-regime. \textbf{e.} Intracavity intensity
for the comb state in (d) showing five solitons ($T_\mathrm{R}$ roundtrip time). \textbf{f.} Zoom into one of the soliton states showing the numerical results for the field real (red) and imaginary part (dark blue). The respective analytical soliton solutions are shown in light blue and orange.}
\end{figure}

\begin{figure}[htbp]
\centering
\includegraphics[width=1\textwidth]{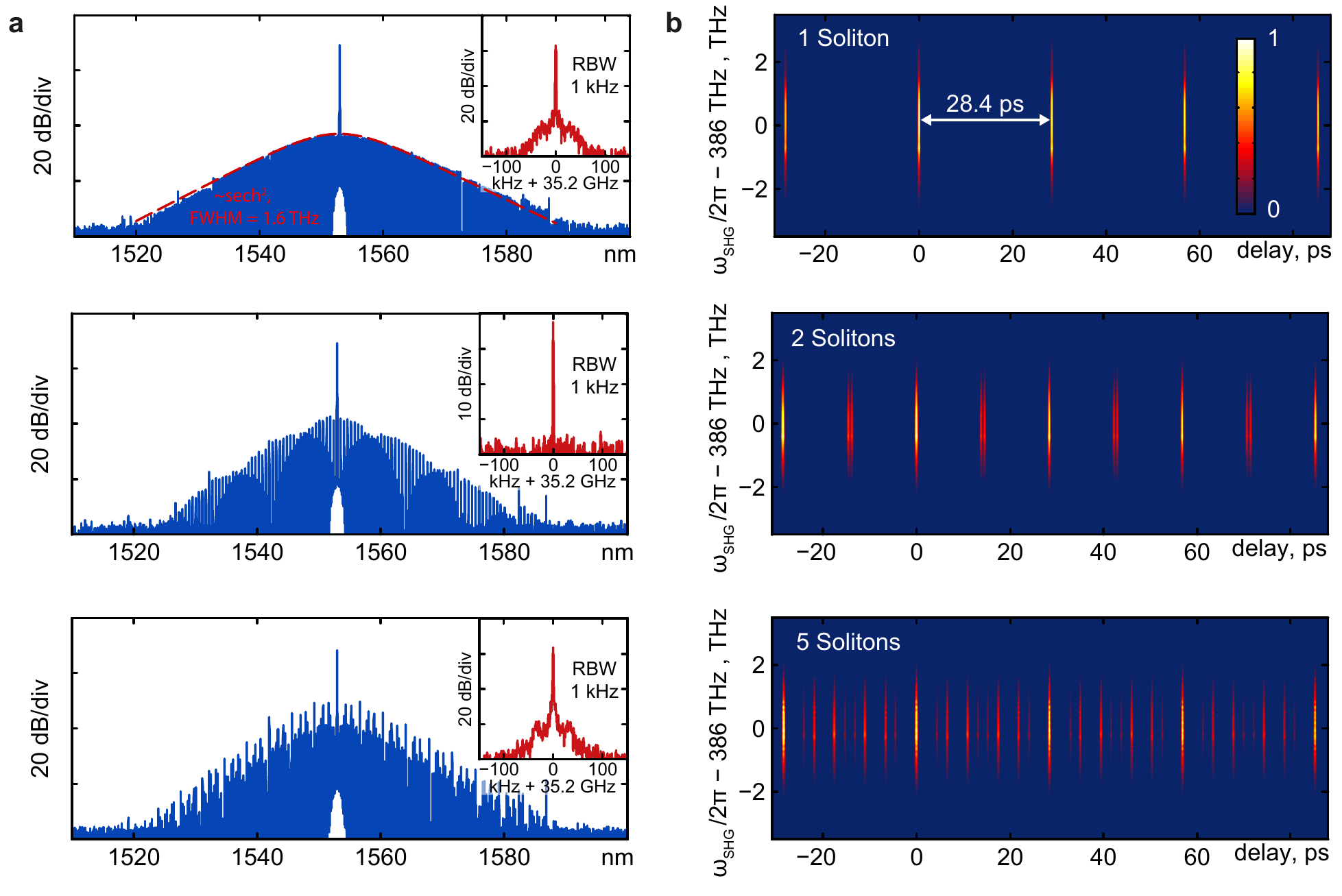}
\caption{\textbf{Experimental demonstration of temporal dissipative soliton states} \textbf{a.}
Optical spectra of three select comb states with one, two and five solitons, respectively.
The insets show the RF beat note, which is resolution bandwidth limited to 1 kHz width in all cases. The red line in the optical spectrum of the one pulse state shows the spectral hyperbolic-secant envelope expected for soliton pulses with a of 3~dB bandwidth of 1.6 THz.
\textbf{b.} FROG traces of the comb states in (a) displaying the signal of the single and multiple pulses. (The FROG setup is shown in Fig.~5b) }
\end{figure}

\begin{figure}[htbp]
\centering
\includegraphics[width=1\textwidth]{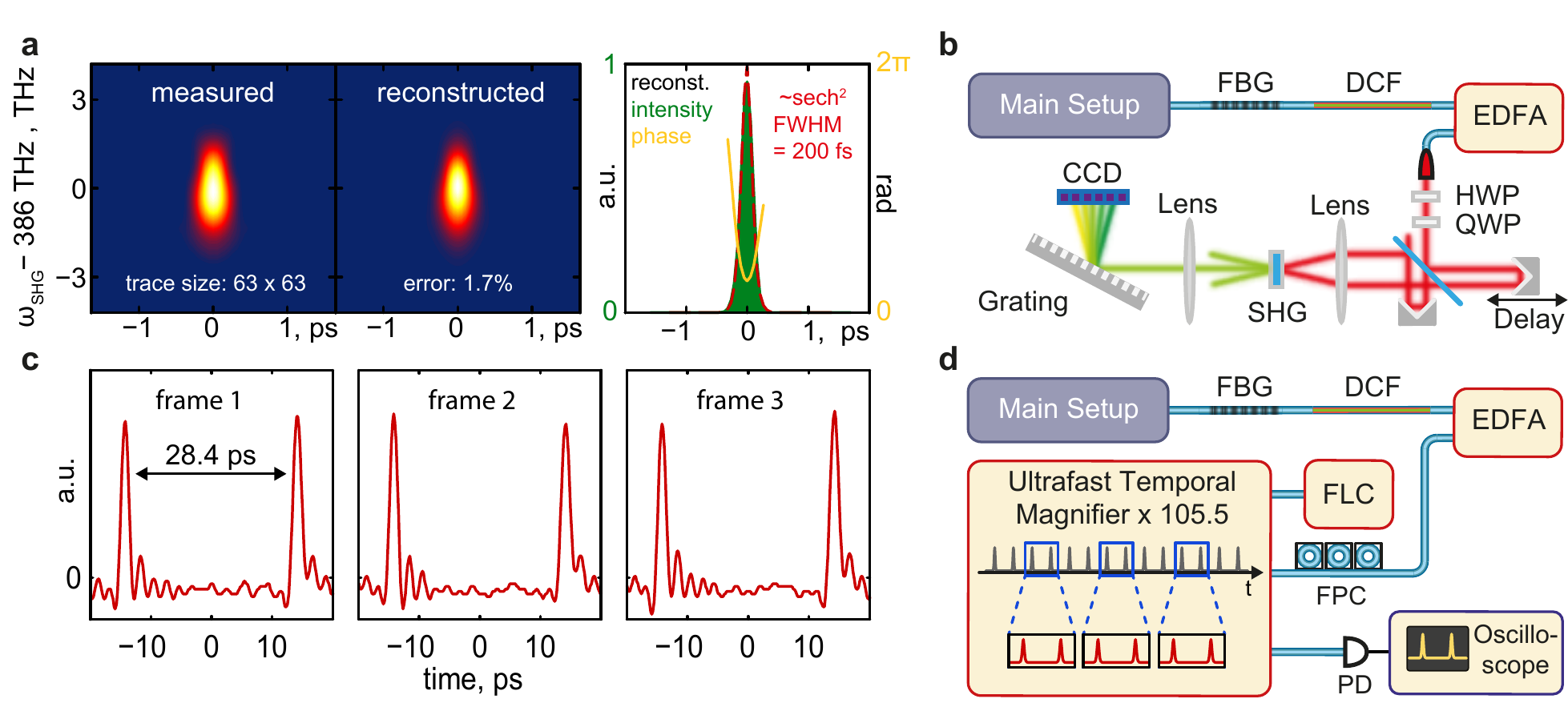}
\caption{\textbf{Temporal characterization of ultra-short pulses} \textbf{a.} Higher resolution experimental FROG trace of a one soliton pulse (left). The reconstruction converges to a FROG error of $\epsilon=1.7$\% in good agreement with the experimental trace (middle).
The reconstruction (right) of intensity and phase yields an estimated pulse duration of 200 fs (FWHM). \textbf{b.} Setup of the FROG experiment. \textbf{c.} Sampled optical intensity of the microresonator output over a duration of $40$~ps. The three measurements (frame 1,2,3) are separated from one another by a duration of $4$~ns corresponding to approx. 140 round-trip times $T_\mathrm{R}=28.4$~ps \textbf{d.} Setup of the intensity sampling experiment including the PicoLuz LLC Temporal Magnifier and a 4~GHz sampling oscilloscope. FBG: fiber-Bragg grating (in transmission), DCF: dispersion compensating fiber, EDFA: erbium-doped fiber amplifier, SHG: second harmonic generation, FLC: fiber laser comb (250MHz rep. rate), PD: photodetector, FPC: fiber polarization controller, HWP: half-wave-plate, QWP: quarter-wave-plate. }
\end{figure}

\clearpage

\section*{\huge Supplementary Information}

\section{Numerical simulation}
When a pump laser with frequency $\omega_p$ is coupled to a high-Q Kerr-nonlinear resonator, a system of nonlinear coupled mode equations \cite{Chembo2010,Chembo2010a,Chembo2010b,Chembo2010d,Maleki2010,Matsko2011a,Herr2012,Chembo2013} can be used to describe the evolution of the mode amplitudes $A_\mu$, which describe the number of photons $|A_\mu|^2$ in the mode with index $\mu$ and resonance frequency $\omega_\mu = \omega_0+D_1 \mu+\tfrac{1}{2} D_2 \mu^2 + \tfrac{1}{6}D_3\mu^3 + ...$ ($D_1=2\pi/T_R$, $D_2$ and $D_3$ corresponding to FSR, second and third order dispersion; fourth and higher order dispersion terms may be introduced in analogous manner). All mode numbers $\mu$ are defined relative to the pumped mode $\mu=0$. The set of coupled mode equations reads:

\begin{align}
\frac{\partial A_\mu}{\partial t}&=-\frac{\kappa}{2} A_\mu+\delta_{\mu_0}\sqrt{\eta\kappa}s_\mathrm{in} e^{-i(\omega_p-\omega_0)t}+ig\!\!\!\!\!\sum_{\mu',\mu'',\mu'''}\!\!\!\!A_{\mu'}A_{\mu''}A^*_{\mu'''}e^{-i(\omega_{\mu'}+\omega_{\mu''}-\omega_{\mu'''}-\omega_\mu)t},\nonumber\\
s_\mathrm{out} &= s_\mathrm{in}-\sqrt{\eta\kappa}\sum A_\mu e^{-i(\omega_\mu-\omega_p)t}.
\end{align}
Here, $\kappa = \kappa_0 + \kappa_{ext}$ denotes the cavity decay rate as a sum of intrinsic decay rate $\kappa_0$ and coupling rate to the waveguide $\kappa_{ext}$, $\eta=\kappa_{ext}/\kappa$ is coupling efficiency ($\eta=1/2$ for critical coupling), and $|s_\mathrm{in,out}| = \sqrt{P_\mathrm{in,out}/\hbar\omega_0}$ denote the amplitudes of the pump and output powers, and $\delta_{\mu 0}$ is Kronecker's delta. The nonlinear coupling coefficient
\begin{align}
g&=\frac{\hbar\omega^2_0 cn_2}{n_0^2V_\mathrm{eff}}.
\end{align}
describes the cubic Kerr-nonlinearity of the system with the refractive index $n_0$, nonlinear refractive index $n_2$, the effective cavity nonlinear volume $V_\mathrm{eff}={\cal A_\mathrm{eff}}L$ (with effective nonlinear mode-area ${\cal A_\mathrm{eff}}$ and circumference of the cavity $L$), the speed of light $c$ and the Planck constant $\hbar$. The summation includes all $\mu',\mu'',\mu'''$ respecting the relation $\mu=\mu'+\mu''-\mu'''$. 

The values of $D_{1}/2\pi=35.2$ GHz, $D_{2}/2\pi=10$ kHz and $D_{3}/2\pi=-130$ Hz were measured following ref \cite{DelHaye2009a} for a resonator similar to the one employed in the present experiments. The mode volume $V_\mathrm{eff}=5.6 \times 10^{-13}~$~m$^3$ was inferred from finite element simulations (such as shown in Fig.1 in the main text). 

The system of coupled mode equations may be rewritten in a dimensionless way, where the explicit time dependence in the nonlinear terms is removed. This is achieved by using the scaling $f=\sqrt{\frac{8g\eta P_\mathrm{in}}{\kappa^2 \hbar \omega_0}}$, $\tau=\kappa t/2$ and phase transformation $a_\mu=A_\mu\sqrt{2g/\kappa}e^{-i(\omega_\mu-\omega_p-\mu D_1)t}$:
\begin{align}
\frac{\partial a_\mu}{\partial \tau}&=-[1+i\zeta_\mu] a_\mu +i\sum_{\mu'\leq \mu''}(2-\delta_{\mu'\mu''}) a_{\mu'} a_{\mu''} a^*_{\mu'+\mu''-\mu}+ \delta_{0\mu} f.
\label{simeqs}
\end{align}
Here $a_\mu$ can be interpreted as the slowly varying  amplitude of the comb modes close to the mode frequency $\omega_\mu$ and $\tau=\kappa t/2$ denotes the normalized time. 

The quantity $\zeta_{\mu}=2(\omega_\mu-\omega_p- \mu D_1)/\kappa$ is a formal measure of detuning defined by the cold resonance frequencies $\omega_\mu$ and an equidistant $D_1$-spaced frequency grid. In the dimensionless form all frequencies, detunings and magnitudes are measured in units of cold cavity resonance linewidth so that $|a_\mu|^2=1$ corresponds to the nonlinear mode shift of half a cold resonance width,  which also corresponds to both single mode bistability and degenerate oscillations thresholds \cite{Chembo2010}.

This set of coupled mode equations (\ref{simeqs}) serves as the basis for the numerical simulations. It is propagated in time using an adaptive stepsize Runge-Kutta integrator. Random vacuum field fluctuation are introduced to seed the initial degenerate four-wave-mixing process. The time of simulation grows cubically with the number of modes taken into account. If $2K$ sidebands and the pump are considered, then the total number of nonlinear terms in all $2K+1$ equations is $\frac{1}{3}(K+1)(8K^2+7K+3)$. 

 As opposed to the modes $A_\mu$, the fields $a_\mu$ correspond to an equidistant frequency grid. Note that amplitude and phase modulation implicitly included in the time dependence of $a_\mu$ include frequency deviations from the equidistant grid and in particular noisy comb states with multiple lines per resonance. For the stationary soliton solutions discussed in the main text and later on here, the amplitudes and phases of $a_\mu$ are constant in time when third and higher order dispersions are neglected. 

Throughout the simulations we neglect thermal effects. We also neglect the frequency dependence of nonlinearity, losses and mode-overlap, interactions with other mode families, and any particularities of the resonator geometry.

In our numerical simulations we also test the influence of the higher order dispersion terms on the formation of soliton states. We varied $D_{3}$ from 0 to $10^{4}$~s$^{-1}$ and found that this does not prevent soliton formation and only affects the pulse repetition rate.


\section{Analytical description of solitons in a microresonators}
To describe the internal field in a nonlinear microresonator the  Nonlinear Schr\"odinger Equation (NLS) may be used when third and higher order dispersion terms are neglected \cite{Matsko2011a}:
\begin{align}
\frac{\partial A}{\partial  t}-i\frac{1}{2} D_2 \frac{\partial^2 A}{\partial \phi^2} - i g |A|^2A
= -\left(\frac{\kappa}{2} +i(\omega_0-\omega_p)\right)A+\sqrt{\frac{\kappa\eta P_\mathrm{in}}{\hbar\omega_0}}.
\label{nls1}
\end{align}
Here $A(\phi,t)=\sum_\mu A_\mu e^{i\mu\phi-i(\omega_\mu-\omega_p) t}$ is the slowly varying field amplitude and $\phi$ is the angular coordinate inside the resonator.
This equation may be formally obtained from the nonlinear equation for the slowly varying amplitude in time domain by using the formal substitution:
\begin{align}
\omega_\mu = \omega_0+D_1 \mu+\frac{1}{2}D_2 \mu^2 = \omega_0+D_1 \mu-\frac{1}{2}D_2 \frac{\partial^2}{\partial \phi^2},
\end{align}
as $\sum_\mu (i\mu)^n A_\mu e^{i\mu\phi-i(\omega_\mu-\omega_p) t}=F.T.\left[\frac{\partial^n}{\partial \phi^n}A(\phi,t)\right]$ 
(details of analogous derivation for a fiber are given in e.g. \cite{Boyd2007}).

Transforming eq. (\ref{nls1}) to its dimensionless form gives:
\begin{align}
i\frac{\partial\Psi}{\partial\tau}+\frac{1}{2}\frac{\partial^{2}\Psi}{\partial \theta^{2}}+|\Psi|^{2}\Psi=(-i+\zeta_{0})\Psi+if.
\label{eq:nls}
\end{align}
Here $\theta=\phi\sqrt{\frac{1}{2d_{2}}}$ is the dimensionless longitudinal coordinate, $\Psi(\tau,\phi)=\sum a_{\mu}(\tau)e^{i\mu\phi}$ is the waveform, and $d_{2}=D_{2}/\kappa$ is the dimensionless dispersion.
Equation (\ref{eq:nls}) is identical to the Lugiato-Lefever equation \cite{Lugiato1987}, where a transversal coordinate is used instead of a longitudinal one in our case. Using the ansatz of a stationary ($\frac{\partial\Psi}{\partial\tau}=0$) soliton on a continuous-wave (cw) background \cite{Wabnitz1993} we find an expression for a single soliton
\begin{align}
 & \Psi=\Psi_{0}+\Psi_{1}\simeq\Psi_{0}+B e^{i\varphi_{0}}{\rm sech}(B\theta),\label{eq:singlesoliton}
\end{align}
where the real number $B$ defines both, width and amplitude of the soliton and $\varphi_{0}$ defines the phase angle. We note that eq. (\ref{eq:singlesoliton}) is not an exact solution of eq. (\ref{eq:nls}), for which exact soliton solutions are known only in the case of zero losses \cite{Barashenkov1996}. 

The constant cw background $\Psi_{0}$ can be found by inserting $\Psi_0$ into eq. (\ref{eq:nls}) as the lowest branch \cite{Barashenkov1996} of the solution of 
\begin{align}
 & (|\Psi_{0}|^{2}-\zeta_{0}+i)\Psi_{0}=if,
\end{align}
which, eventually, results when $\zeta_{0}>\sqrt{3}$ (bistability criterion) and large enough detunings $f^2<\frac{2}{27}\zeta_0(\zeta_0^2+9)$ in: 
\begin{align}
|\Psi_0|^2&=\frac{2}{3}\zeta_0-\frac{2}{3}\sqrt{\zeta_0^2-3}\cosh\!\left(\frac{1}{3}\,\,\mathrm{arcosh}\!\left(\frac{2\zeta_0^2+18\zeta_0-27f^2}{2(\zeta_0^2-3)^{2/3}}\right)\right),\nonumber\\
\Psi_0&=\frac{if}{|\Psi_0|^2-\zeta_0+i}\simeq\frac{f}{\zeta_{0}^{2}}-i\frac{f}{\zeta_{0}}.
\label{eq:Psi0}
\end{align}

The soliton component $\Psi_{1}$ in eq. (\ref{eq:singlesoliton}) is approximated by the bright soliton solution of the undriven, undamped (ordinary) NLS, which is the limit case for $\zeta_{0}\gg1$.

The parameters $B$ and $\varphi_{0}$ can be derived based on general conditions for the soliton attractor \cite{Wabnitz1993}, which yields 
\begin{align}
B&\simeq\sqrt{2\zeta_{0}},
\label{eq:B}
\end{align}
\begin{align}
\cos(\varphi_{0})& \simeq\frac{\sqrt{8\zeta_{0}}}{\pi f}. \label{solparms}
\end{align}

Based on eqs. (\ref{eq:Psi0},\ref{eq:B}) we can estimate the ratio $R$ of soliton peak power to cw pump background:
\begin{align}
R=\frac{|B|^2}{|\Psi_0|^2}=\frac{2\zeta_0^3}{f^2}.
\end{align}
For a maximal detuning $\zeta_0 = \zeta_0^\mathrm{max}$ (see eq. (\ref{eq:lim1}) below) we find:
\begin{align}
R_\mathrm{max}=\frac{\pi^6 f^4}{2^8}= \left(\frac{\pi^3 g\eta P_\mathrm{in}}{2\kappa^2 \hbar \omega_0}\right)^2.
\end{align}

Extending eq. (\ref{eq:singlesoliton}) to the case of multiple solitons inside the resonator gives 
\begin{align}
 & \Psi(\phi)\simeq \underbrace{\Psi_{0}}_{C_1}+\underbrace{\left(\frac{4\zeta_{0}}{\pi f}+i\sqrt{2\zeta_{0}-\frac{16\zeta_{0}^{2}}{\pi^{2}f^{2}}}\right)}_{C_2}\sum_{j=1}^{N}\,\mathrm{sech}(\sqrt{\frac{\zeta_{0}}{d_{2}}}(\phi-\phi_{j})).\label{waveform}
\end{align}

It was shown in \cite{Wabnitz1993} that if a pair of solitons in a train is separated by a distance $\phi_{j+1}-\phi_{j} \gtrsim (8/B) \sqrt{2d_2}$ the pair of solitons does not interact. This puts a possible limit $N_\mathrm{max} < \frac{2\pi}{8}\sqrt{\zeta_0/d_2}$ of a maximum number of stationary solitons in the resonator and consequently the maximum number of ``steps'' in intracavity power and transmission. Assuming that soliton can only emerge for $\zeta_0>\sqrt{3}$ (bistability criterion), we find $N_\mathrm{max} \approx \sqrt{\kappa/D_2}$, which remarkably coincides with the distance between the first generated primary sidebands $\mu_\mathrm{th, min}$ in the process of comb generation \cite{Herr2012}. 

\section{Limit conditions for solitons in microresonators}
\label{limitssection}
By substituting $|h|=|f|/\sqrt{2\zeta_{0}^{3}}$, $\gamma=1/\zeta_{0}$, $\tilde\theta=\sqrt{2\zeta_0}\theta$, $\Psi=\sqrt{2\zeta_0}\tilde\Psi$ and changing the phase of the pump, (\ref{eq:nls}) is transformed to the damped driven NLS equation for
the stationary case:
\begin{align}
\frac{\partial^{2}\tilde\Psi}{\partial \tilde\theta^{2}}+2|\tilde\Psi|^{2}\tilde\Psi-\tilde\Psi=-i\gamma\tilde\Psi-h,\label{eq:nls-1}
\end{align}
which was analyzed for infinite boundary conditions in \cite{Barashenkov1996}. In particular the condition for the soliton existence
$h>2\gamma/\pi$ transforms into:
\begin{align}
\zeta_{0}^\mathrm{max}=\pi^{2}f^{2}/8.\label{eq:lim1}
\end{align}
Eq. (\ref{eq:lim1}) can also be found from the requirement that the right part  in equation (\ref{solparms}) must be smaller than unity.  
In \cite{Barashenkov1996} it was further shown analytically that the boundaries separating the regimes of existence of solitons (as described
in the main text) are defined by characteristic curves for $\tilde\Psi_{0}$ in (\ref{eq:nls-1}). In our case this translates into 
\begin{align}
|\Psi|_{\pm}^{2}=\frac{2}{3}\zeta_{0}\pm\frac{1}{3}\sqrt{\zeta_{0}^{2}-3}
\end{align}
Numerical simulations for our system with periodic boundary conditions show that all these limits remain valid with very good quantitative agreement for a sufficiently large number  $K\gg \frac{2}{\pi}\sqrt{\frac{\zeta_0^\mathrm{max}}{d_2}}=f\mu_\mathrm{th, min}/\sqrt{2}$ of modes (typically a few hundred).

\section{Analytical description of steps in the intracavity power}
The height of steps in the intracavity power can be found by averaging the waveform amplitude (eq. \ref{waveform}) squared over one roundtrip for different numbers $N$ of solitons:
\begin{align}
\overline{|\Psi|^{2}}&=|\Psi_{0}|^{2}+N\cdot\xi(K)\frac{1}{2\pi}\int_{0}^{2\pi}(\Psi_{1}^{2}+\Psi_{0}\Psi_{1}^{*}+\Psi_{1}\Psi_{0}^{*}) d\phi\nonumber\\
&=|\Psi_{0}|^{2}+N\sqrt{2d_2}(\Psi_0'\cos\phi_0+\Psi_0''\sin\phi_0+\sqrt{2\zeta_0}/\pi)\simeq \frac{f^2}{\zeta_0^2}+N\xi(K)\frac{2}{\pi}\sqrt{d_2\zeta_0}.
\end{align}
As shown in Fig.~3 in the main manuscript, this approach also describes the laser tuning dependence of the step height. When comparing to the numerical simulations with a rather low mode number, we use a correction factor $\xi(K)$ of order unity $(\xi(K)\simeq1.3$ in the case of 101 simulated modes). For a higher number of simulated modes (e.g. $501$) this correction is not required.

\section{Optical spectrum and temporal width of solitons in a microresonator}
The optical spectrum $\Psi(\mu)$ of the soliton has the same hyperbolic secant form as the time domain waveform. Mathematically this corresponds to the Fourier transform of a hyperbolic secant being again a hyperbolic secant:
\begin{align}
\Psi(\mu)=\mathrm{F.T.}\left[\sqrt{2\zeta_{0}}\,\mathrm{sech}\left(\sqrt{\frac{\zeta_{0}}{d_{2}}}\phi\right)\right]=\sqrt{d_{2}/2}\,\mathrm{sech}\left(\frac{\pi\mu}{2}\sqrt{\frac{d_{2}}{\zeta_{0}}}\right).
\end{align}
Using the relation for the optical frequency $\omega = \omega_p + \mu D_1$ and the time $t=\tfrac{\phi}{2\pi} T_\mathrm{R}=\phi/D_1$ spectral envelopes and the soliton waveform can be rewritten:
\begin{align}
\Psi(\omega-\omega_{p}) = \sqrt{d_{2}/2}\,\mathrm{sech}((\omega-\omega_{p})/\Delta\omega) \ \  \mathrm{with} \ \Delta\omega=\frac{2D_{1}}{\pi}\sqrt{\frac{\zeta_{0}}{d_{2}}}.
\end{align}
and 
\begin{align}
\Psi(t)=\sqrt{2\zeta_{0}}\,\mathrm{sech}(t/\Delta t),\ \ 
\mathrm{with} \ \Delta t=\frac{1}{D_{1}}\sqrt{\frac{d_{2}}{\zeta_{0}}}.
\end{align}
The minimal achievable soliton duration can be found by using $\zeta_0^\mathrm{max}$ (eq. \ref{eq:lim1}) in the above equation for $\Delta t$:
\begin{align}
 \Delta t_\mathrm{min}=\frac{1}{\pi D_{1}}\sqrt{\frac{\kappa D_2  n_0^2 V_\mathrm{eff}}{\eta P_\mathrm{in} \omega_0 c n_2}}.
\end{align}
This equation can be recast in form of the group velocity dispersion $\beta_2 = \tfrac{-n_0}{c}D_2/D_1^2$, the nonlinear parameter $\gamma=\tfrac{\omega}{c}\tfrac{n_2}{\cal A_\mathrm{eff}}$ (for simplicity we assume critical coupling $\eta=1/2$ and on resonant pumping):
\begin{align}
 \Delta t_\mathrm{min}=\frac{2}{\sqrt{\pi}}\sqrt{\frac{-\beta_2}{\gamma \cal{F} P_\mathrm{in}}},
\end{align}
where denotes the finesse ${\cal F}=\tfrac{D_1}{\kappa}$ of the cavity. Note that the values $\Delta \omega$ and $\Delta t$ need to be multiplied by a factor of $2\,\mathrm{arccosh}(\sqrt{2})=1.763$ to yield the FWHM of the sech$^2$-shaped power spectrum and pulse intensity, respectively. 

For the time bandwidth product (TBP) we find $\Delta t\Delta\omega=2/\pi$ or, when considering the FWHM of spectral and temporal power (in units of Hz and s), $\mathrm{TBP}=0.315$.

In the case of $N$ multiple solitons inside the cavity there is a more structured optical spectrum $\Psi_{N}$, resulting from interference
of single soliton spectra $\Psi_{j}(\mu)$, where the relative phases of these spectra are determined by the positions $\phi_{j}$ of individual solitons: $\Psi_{N}(\mu)=\Psi(\mu)\sum_{j=1}^{N}\mathrm{e^{i\mu\phi_{j}}}$. The line-to-line variations can be high, however, the overall averaged spectrum still follows the single soliton shape and is proportional to 
$\sqrt{N}\Psi(\mu)$. 

\section{Laser tuning method to achieve soliton states}
Experimentally the soliton states in the MgF$_{2}$ resonator can not be stably reached by slowly (manually) tuning into the soliton state. The obstacle lies in the temperature drop the resonator experiences when transiting from the upper branch state (high intracavity power) to the lower branch soliton state (lower intracavity power). This sudden temperature drop leads to a blue-shift of the resonance frequency and a loss of the soliton state. On the other hand, when tuning very quickly into the soliton state the resonator is still cold and its subsequent heating will again lead to a loss of the soliton state. 

\begin{figure}
\centering
\includegraphics[width=0.6\textwidth]{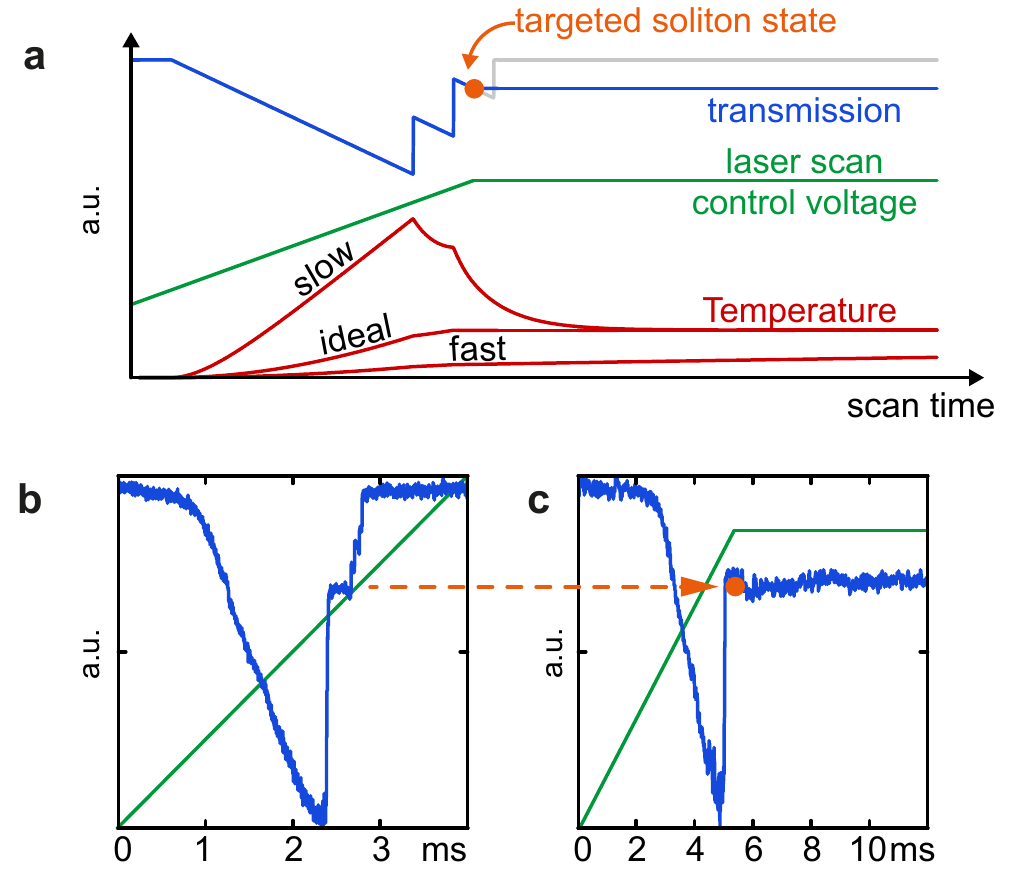}
\caption{\textbf{Laser tuning method to achieve soliton states.} \textbf{a.} Illustration of the laser tuning method, where a laser frequency scan (green) is performed that stops when the targeted comb state, marked by a orange dot in the corresponding transmission signal (the grey line illustrates the signal that would have been observed if the scan had been continued). The system remains stably in this state when the appropriate scan speed is chosen. In this ideal scenario the temperature (which starts increasing as soon as light is coupled to the resonator) reaches the steady-state equilibrium temperature of the targeted state when the system has reached this state via laser detuning. If the laser scan is performed too slow (fast) then the resulting temperature will be too high (low) and destabilize the system. 
\textbf{b.} Experimental laser scan over a resonance, showing a pronounced step followed by multiple smaller steps. \textbf{c.} Demonstration of the adaptive scanning method. The laser scan is stopped after the transition to the soliton regime. The appropriate choice of scan speed allows the system to operate stably in a soliton state. The coupled pump power is 30~mW.}
\label{fig:laserTuningMethod}
\end{figure}

We solve this problem by tuning into the soliton state with an ideal, intermediate tuning speed, such that the resonator reaches the soliton state in a thermal equilibrium, that is, neither too hot nor too cold. This method is illustrated in Fig.~\ref{fig:laserTuningMethod}. Practically this is achieved by programming an electronic laser frequency ramp signal defined by the three parameters laser tuning speed, and laser start and end wavelength. This signal is used to control the piezo tuning of the fiber laser. The laser frequency ramp is performed in the direction of decreasing optical frequency in order to first follow the upper branch before jumping onto the lower branch. Once found for a particular resonator, the parameters do not need to be changed and allow reliable generation of soliton states at the push of a button. In contrast to fiber cavity experiments, the soliton pulses form spontaneously without the need for external stimulation. The number of solitons formed can be controlled by the pump laser detuning. Once generated the solitons remain stable for hours until the pump laser is switched off without the need for active feedback on neither the resonator nor the pump laser. The stability of the soliton states is discussed in the next Section.
\section{Self-stability of soliton states}
As it is discussed in the main text the soliton states are achieved when the pump laser is detuned to the lower, effectively red detuned branch (cf. Fig.~2e in the main text). The generated solitons remain stable without any active external feed back applied to the system. This is remarkable and indeed surprising as operating on the lower branch (effectively red detuned) is usually thermally unstable\cite{Ilchenko1992a,Carmon2004} and would require active stabilization techniques. The thermal instability of the lower branch of the Kerr-bistability curve can be explained by the negative slope (decreasing intracavity power for increased laser detuning $\zeta_0$). 

\begin{figure}
\centering
\includegraphics[width=0.75\textwidth]{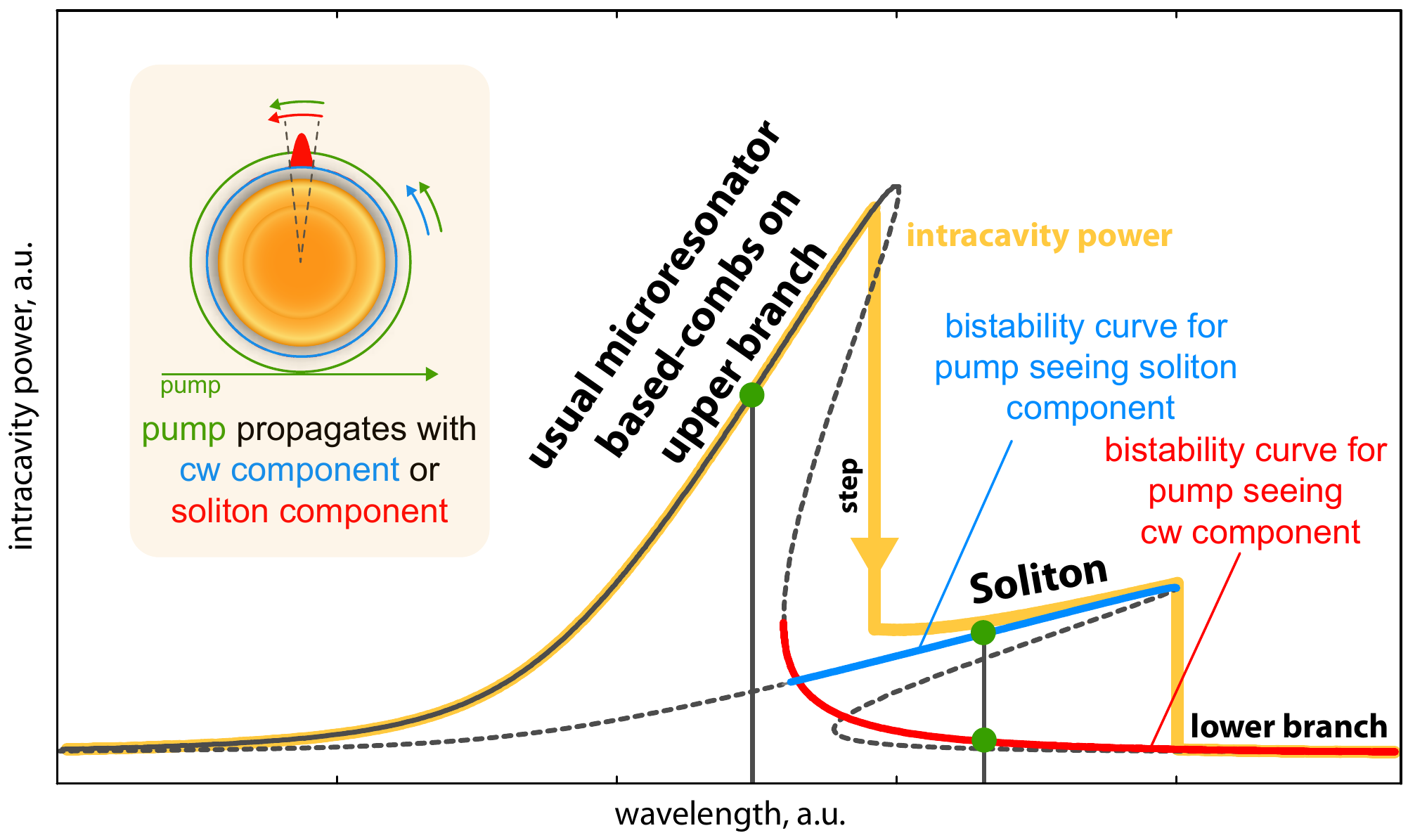}
\caption{\textbf{Thermal self-stability of soliton states:} This figure is a more detailed version of Fig.~1c in the main text. The higher level of detail is required to explain the self-stability of soliton states. When tuning into the resonance  with increasing wavelength (decreasing optical frequency) the intracavity power is described by the upper branch of the bistability curve. After the transition to a soliton state the major fraction of the pump light is described by the lower branch bistability curve (red). The fraction of the pump light that propagates with the soliton inside the microresonator (cf. inset) experiences a larger phase shift and is effectively blue detuned on the upper branch of another bistability curve (blue). The shape of the latter bistability curve depends on the number of solitons present in the cavity and their peak power as discussed in the text. The resulting intracavity power can be inferred by adding the bistability curves all fractions of the pump light resulting in the yellow curve.  A positive slope in the combined average intracavity power implies thermal stability of the system.}
\label{fig:DualBistabCurve}
\end{figure}

In the presence of solitons, however, the fraction of the pump light that propagates at a similar velocity (The difference between phase and group velocity will result in an effectively longer soliton pulse. The presented reasoning remains valid). together with the high intensity solitons inside the resonator, experiences a much larger phase shift in one cavity roundtrip, compared to the fraction seeing only the cw component. One may interpret the situation in terms of two superimposed bistability curves; one bistability curve for the fraction of the pump light overlapping with the cw component and one bistability curve for the fraction of the pump light overlapping with the solitons. Fig.~\ref{fig:DualBistabCurve} illustrates the respective bistability curves of the intracavity average power for the case of one soliton. The bistability curve resulting from the soliton is much lower in height as the spatial length of a soliton is small compared to the roundtrip length of the resonator. Due to the higher peak power (when compared to the intracavity cw component) the bistability curve resulting from the soliton extends to larger detunings. While the main portion of the pump light is effectively red detuned (lower branch) the small portion overlapping with the soliton inside the resonator is effectively blue detuned (upper branch). Importantly, the resulting combined average intracavity power (the sum of the two bistability curves) has a positive slope (increasing intracavity power for increasing laser wavelength), which is responsible for the effective thermal self-stability. The resulting positive slope can indeed be evidenced for various number of solitons in the experiment (cf. main text Fig.~2a, a negative transmission slope corresponds to a positive slope in average intracavity power), as well as in the numerical simulation (blue curve, main text Fig.~3a). It is important to note that a combined positive slope requires that the positive slope (induced by the soliton) dominates over the negative slope (related to the intracavity cw component). This puts a limit on the maximum length of the cavity: For long cavities, such as fiber cavities, the intracavity power is dominated by the cw component and active stabilization\cite{Leo2010}) of the system is required. This explains why self-stability of the solitons is only observed in microresonators. Finally, we note that while the intracavity power in the soliton state is dominated by the soliton component, the PDH signal is dominated by the large fraction of the pump light that is effectively red detuned (Experimentally this can be seen by the only small changes to the PDH signal in the red detuned regime when transmission steps occur, cf. main text Fig.~2e.
\section{Frequency resolved optical gating (FROG) vs. the auto-correlation technique}
\begin{figure}
\centering
\includegraphics[width=0.8\textwidth]{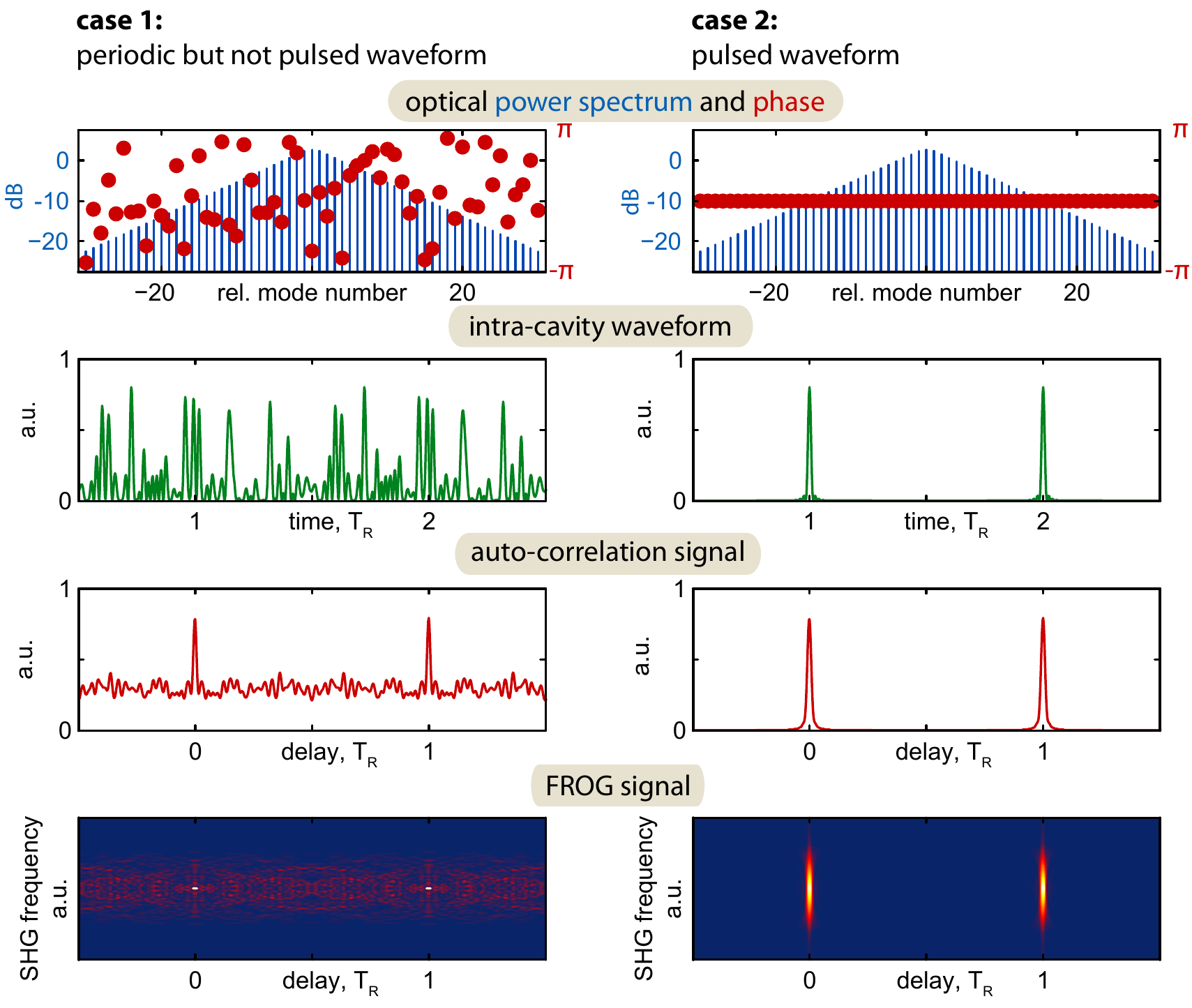}
\label{fig:timedomainmethod}
\caption{\textbf{Autocorrelation vs. frequency resolved optical gating} The left column (case 1), shows the situation of a comb spectrum with phases which are constant in time, but arbitrary and not such that a pulse forms inside the microresonator ($T_\mathrm{R}$ denotes the light roundtrip time of the microresonator).  The right column (case 2) corresponds to the situation, where a single pulse circulates inside the microresonator. The expected auto-correlation and FROG signals are shown for both cases. Note that the intra-cavity waveforms as well as the autocorrelation and FROG signals have been rescaled to the same peak values. In the pulsed case 2 the reached peak intensities in the intracavity waveform and consequently in the auto-correlation and FROG signal are much higher than in the not pulsed case 1.}
\label{fig_SI_FROGvsAutocorr}
\end{figure}

In microresonator based systems the proof of a truly pulsed time-domain waveform inside the microresonator (and pulses coupled out directly form a the microresonator) is challenging.  The difficulty lies in distinguishing the truly pulsed scenario (where all phases equal such that a pulse forms) from the case where all phases between comb lines are constant in time but arbitrary, which results in a periodic, modulated time domain output, which however is not truly pulsed, eg. \cite{Ferdous2011, Papp2011a}. Figure~\ref{fig_SI_FROGvsAutocorr} compares the two cases (not pulsed and pulsed), based on simulated data. While the FROG measurement can clearly distinguish between the the two cases, a simple auto-correlation measurement can not. Indeed, the auto-correlation trace can exhibits narrow spikes, that are easily confused with pulses, even in the case where no pulses are present in the cavity. This applies in particular to experimental data, where detector noise and background further complicate reliable analysis of auto-correlation data. A detailed understanding and discussion of the background in experimental auto-correlation traces is essential to correctly interpret the results. 
\section{Soliton mode-locking in lasers vs. soliton formation in microresonators}
\label{comparison}
This section compares soliton formation in microresonators with soliton mode-locking in lasers where a saturable absorber is necessary for soliton stability: \\

{\bf Soliton mode-locking in lasers:} Generally mode-locking requires a pulse shaping mechanism, which can be achieved in different ways for example via a fast saturable absorber that forms the circulating intensity inside the laser cavity into a pulse\cite{Haus2000a}. Here, the shortest achievable pulse duration is limited by the relaxation time of the fast saturable absorber. Another mode-locking mechanism is soliton mode-locking, where the pulse shaping mechanism is provided by the formation of solitons in the presence of negative group velocity dispersion and self-phase modulation via the cavity’s non-linearity. This method is widely used and well understood in the context of mode-locked laser and ultra-short optical pulse generation\cite{Kaertner1996,Kaertner1998}. While the pulse shaping does not rely on the effect of a saturable absorber, it has been shown theoretically and experimentally that soliton mode-locked lasers still require a saturable absorber, which ensures the stability of the soliton against the growth of a narrow-bandwidth cw background\cite{Kaertner1996,Kaertner1998}. This cw background arises from the interaction and reshaping of the soliton in the laser cavity and subsequently experiences a larger gain as the soliton, which due to its broadband spectral nature falls into the outer, lower gain parts of the spectral laser gain window. It is important to note that in the case of soliton mode-locking the relaxation time of the saturable absorber does not limit the achievable pulse duration; it merely ensures the suppression of the continuum on intermediate timescales\cite{Jung1995}.
\\
\\
{\bf Soliton formation in microresonators:} Soliton formation in microresonators is similar to soliton mode-locking in lasers. As in the case of a soliton-mode locked laser solitons are formed due to a balance between cavity nonlinearity and self-phase modulation. However, while microresonators are driven by a continuous wave pump laser they are not lasers. The conversion of the continuous pump laser light into other frequency components and the amplification of the newly generated frequency components rely exclusively on the parametric gain due to the Kerr-nonlinearity of the resonator material. The cw pump laser coincides directly with a spectral component of the solition. Importantly, a saturable absorber is not required for the stability of the solitons as detailed below: Mathematically the coherently driven, damped Kerr-nonlinear microresonator is described by the Lugiato-Lefever equation\cite{Lugiato1987}, which is identical to a damped, driven nonlinear Schrödinger equation. Dissipative temporal cavity solitons, superimposed onto a weak continuous wave background, have been proven to exist as stable mathematical solutions to this equation\cite{Wabnitz1993}. Due to the cavity loss, these solitons are dissipative in their nature and their persistence requires a source of energy for replenishment. The latter is provided by continuously and coherently driving the cavity. \\

The continuous wave background on which the solitons exist in the case of a microresonator is very different from the detrimental cw background in a soliton mode-locked lasers. It is a coherent internal field originating from the pump laser. It is not a narrow bandwidth low intensity cw background pulse produced by perturbation of the soliton. As opposed to a spectrally limited but continuous laser gain medium (continuous in the sense that it can amplify any frequency component within the gain bandwidth) the parametric gain profile is highly frequency selective (as energy conservation needs to be fulfilled in the frequency conversion processes). Moreover, the parametric gain profile depends on the light frequencies and intensities present in the cavity and relies on the phase coherent interaction between all these light frequencies. While it cannot replace a stringent mathematical stability analysis as mentioned above, this illustrates that the growth of a destabilizing cw background is generally not supported in a microresonator. Hence, stable soliton formation in a microresonator does not require a saturable absorber and the solitons are well described by the Lugiato-Lefever equation. This also is in perfect agreement with our experiments, which reveal the generation of stable solitons in optical microresonators.

\section{Spectral Broadening}
\label{broadening}
\begin{figure}
\centering
\includegraphics[width=1\textwidth]{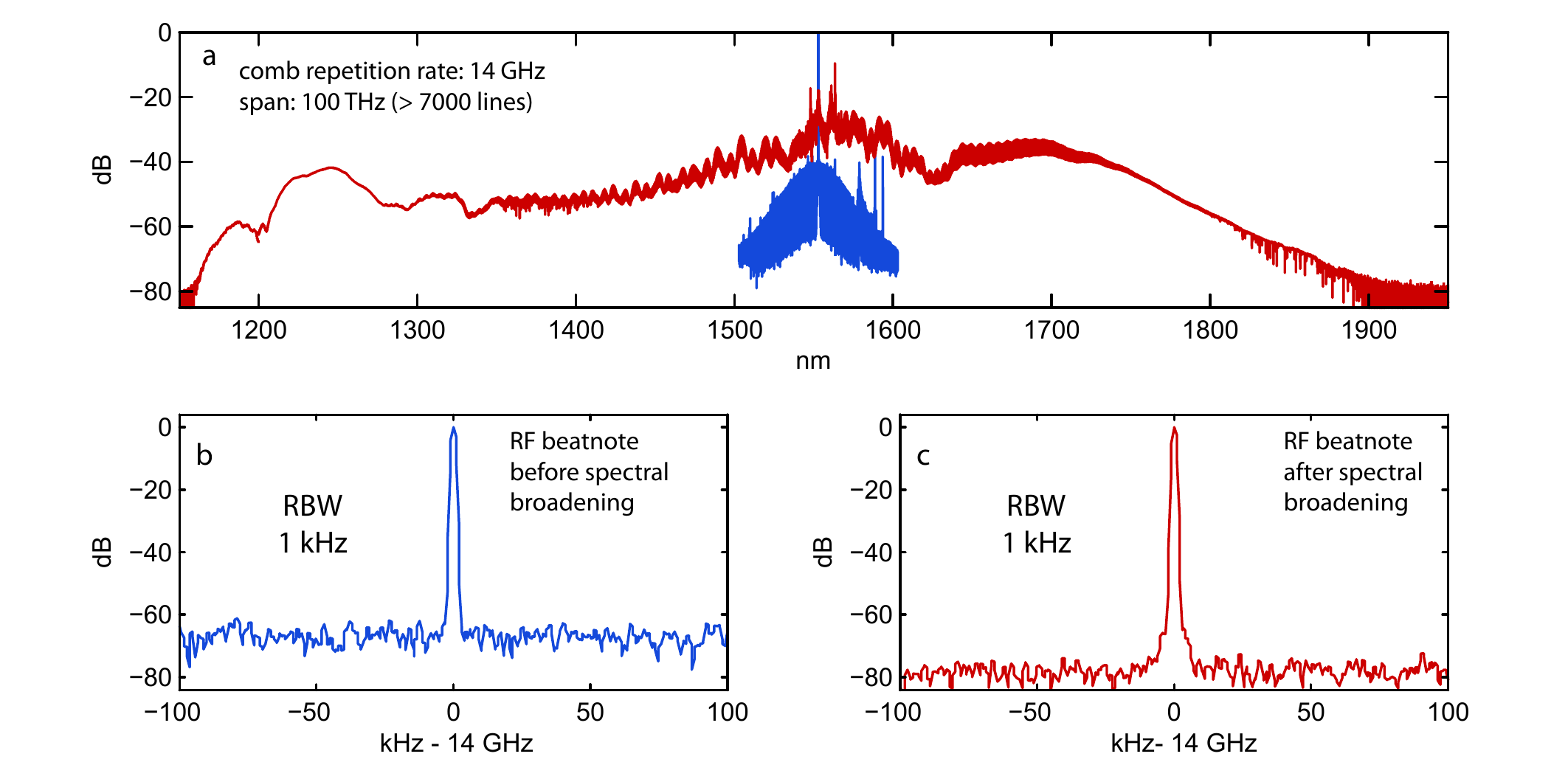}
\label{fig:broadening}
\caption{\textbf{External broadening of ultra-short pulses (estimated 112~fs) from a microresonator:} \textbf{a} Optical single soliton spectrum (blue) generated in a 14.1~GHz MgF$_2$ resonator with 80~kHz resonance width. The spectrum is amplified and broadened to a supercontinuum (red) in a highly-nonlinear fiber. \textbf{b,c} Radio-frequency (RF) repetititon rate beatnote observed when detecting the optical spectra on a fast photodetector before and after spectral broadening (resolution bandwidth 1~kHz). 
}
\end{figure}

Self-referencing via e.g. \textit{f-2f} or \textit{2f-3f} interferometry \cite{Telle1999,Cundiff2003} is an important future technical milestone for microresonator based combs (including soliton based combs). Necessary for these self-referencing schemes is a minimal optical comb bandwidth of an octave (\textit{f-2f}) or two thirds of an octave (\textit{2f-3f}). So far however, self-referencing of microresonator combs has not been possible as no system was capable of generating sufficiently broad spectra while maintaining the low-noise level required for metrology operation. The discovery of soliton formation in microresonators and the generation of ultra-short optical pulses enables spectral broadening using techniques that have been developed for conventional mode-locked lasers. Here we demonstrate in a first proof-of-concept experiment external broadening of a soliton based microresonator frequency comb to a broadband spectrum. Note that we do not employ the same resonator as in the main text but a larger MgF$_2$ resonator with an FSR of only 14.1~GHz. Based on the spectral envelope the pulse duration in the one soliton state is estimated to 112~fs. The pulses are amplified in an erbium doped fiber amplifier to approximately 3.2~W average power. The amplified and dispersion-compensated pulses are sent into 2~m of highly-nonlinear fiber. The achieved spectral bandwidth is close to two thirds of an octave as required for self-referencing (Fig.~\ref{fig:broadening}). No indication of added noise in the RF beatnote is observed.  A more careful analysis of coherence\cite{Coen2002,Dudley2006} in the broadened spectrum is beyond the present scope and subject of future work. The presented results show that external broadening techniques can in principle be applied to microresonator based combs and open a viable route towards future self-referencing of microresonator based combs.

\newpage


\clearpage

\renewcommand{\emph}{\textit}
\bibliographystyle{naturemag}
\bibliography{THBib}

\end{document}